\documentclass{mem}
\usepackage{natbib}\usepackage{txfonts}\usepackage{balance}
\usepackage{graphicx}
\usepackage[a4paper]{hyperref}
\idline{75}{282}
\begin{document}
\def\teff{$T\rm_{eff }$}
\def\kms{$\mathrm {km s}^{-1}$}
\def\ls{{_<\atop^{\sim}}}
\def\gs{{_>\atop^{\sim}}}
\def\cgs{ ${\rm erg~cm}^{-2}~{\rm s}^{-1}$ } 

\title{
Unveiling obscured accretion
}

   \subtitle{}

\author{
F. \,Fiore \& the HELLAS2XMM collaboration
          }

  \offprints{F. Fiore}

\institute{
Istituto Nazionale di Astrofisica --
Osservatorio Astronomico di Roma, via Frascati 33
I-00040 Monteporzio, Italy
\email{fiore@mporzio.astro.it}
}

\authorrunning{Fiore}

\titlerunning{Unveiling obscured accretion}

\abstract{ We present the latest determination of the X-ray (2-10 keV)
AGN luminosity function accounting for the selection effect due to
X-ray absorption.  The main results are: 1) the inclusion of obscured
AGN confirms the AGN differential luminosity evolution, but makes it
less extreme than what is found selecting unobscured AGN in soft
X-rays, and more similar to a pure luminosity evolution; 2) significant
correlations are found between the fraction of obscured sources, the
luminosity and the redshift, this fraction increasing toward both low
AGN luminosities and high redshifts. We discuss our findings in a
scenario for the formation and evolution of the structure in the
Universe where the bulk of nuclear activity is produced at
z$\sim1-2$. At the same redshifts also the star-formation rate reaches
a maximum, and this age can therefore be regarded as the "golden age" for
nuclear and galaxy activity.
We discuss the current observational limits of this program and the
improvements needed to obtain an unbiased census of the AGN and
super-massive black hole (SMBH) population.  }

\maketitle{}

\section{Introduction}

Active Galactic Nuclei are not only witnesses of the phases of galaxy
formation and/or assembly, but most likely among the leading actors.
Indeed, two seminal discoveries indicate tight links and feedbacks
between SMBH, nuclear activity and galaxy
evolution. The first is the discovery of SMBH at the center of most
nearby bulge dominated galaxies, and, in particular, the steep and
tight correlation between their masses and galaxy bulge properties
(see e.g. Ferrarese \& Merrit 2000, Gebhardt et al. 2000, Marconi \&
Hunt 2003).  The second discovery was originally due to the first deep
X-ray surveys performed by ROSAT at the beginning of the 90'. They
showed that the evolution of AGN is luminosity dependent. On average,
the activity of Seyfert like objects rises up to z$\approx1$ and then
decreases, while QSO activity rises smoothly up to z=2-3 or even
further (Miyaji et al. 2000 and references therein).  The former
recalls the evolution of star-forming galaxies, while the latter recals
the evolution of massive spheroids (Franceschini et al.
1999). However, soft X-ray surveys are biased against obscured
sources, which, on the other hand, are very common in the local
Universe. Indeed, the first imaging surveys above 2 keV obtained with
ASCA and BeppoSAX found the first size-able samples of highly obscured
AGN at z$>0.1$ and confirm, at least qualitatively, the predictions of
standard AGN synthesis models for the Cosmic X-ray Background, CXB
(Fiore et al. 1999, Akiyama et al. 2000, La Franca et al. 2002, Ueda
et al. 2003).  Chandra and XMM-Newton surveys confirm and expand this
picture. On one side they resolved nearly 100\% of the CXB below 2 keV
(Giacconi et al. 2002, Brandt et al. 2001) and confirm the strong
luminosity dependent density evolution of soft X-ray sources (Hasinger
2003, 2005).  On the other side deep and large area surveys up to 10
keV clearly showed that AGN activity spans a range of optical to
near--infrared properties much greater than it was thought based on
optically and soft X--ray selected AGNs (Fiore et al. 2003, Koekemoer
et al. 2004, Barger et al. 2005).  The main reason is that hard X--ray
selection provides a more complete and direct view of AGN activity,
being less biased than optical or soft X--ray selection against
obscured sources.  For example, 2-10 keV surveys pick up AGN
relatively bright in X-rays but with extremely faint optical
counterparts.  The majority of these sources have been identified as
highly obscured, high luminosity AGN at z$\gs1$, the so called type 2
QSOs ($\sim20$ in the HELLAS2XMM survey, 3/4 confirmed through optical
spectroscopy, Fiore et al. 2003, Mignoli et al. 2004, Maiolino et
al. 2006, Cocchia et al. 2006, about 30 in the CDFS+CDFN, half of
which confirmed through optical spectroscopy; a dozen in the CLASXS
survey, Barger et al. 2005). At the opposite of the X-ray to optical
flux ratio distribution, Chandra and XMM-Newton hard X-ray surveys
discovered moderately obscured sources with AGN luminosity, in
otherwise inactive, optically bright, early type galaxies (named
X--ray bright, optically normal galaxies, XBONGs, Fiore et al. 2000,
Comastri et al. 2002).

In this paper we discuss how the inclusion of obscured AGN affects the
determination of the AGN luminosity function, and discuss our findings
in the framework of semi-analytical models for the formation and
evolution of the structure in the Universe.

\section{The evolution of hard X-ray selected sources}

\begin{figure}[t!]
\resizebox{\hsize}{!}{\includegraphics[clip=true]{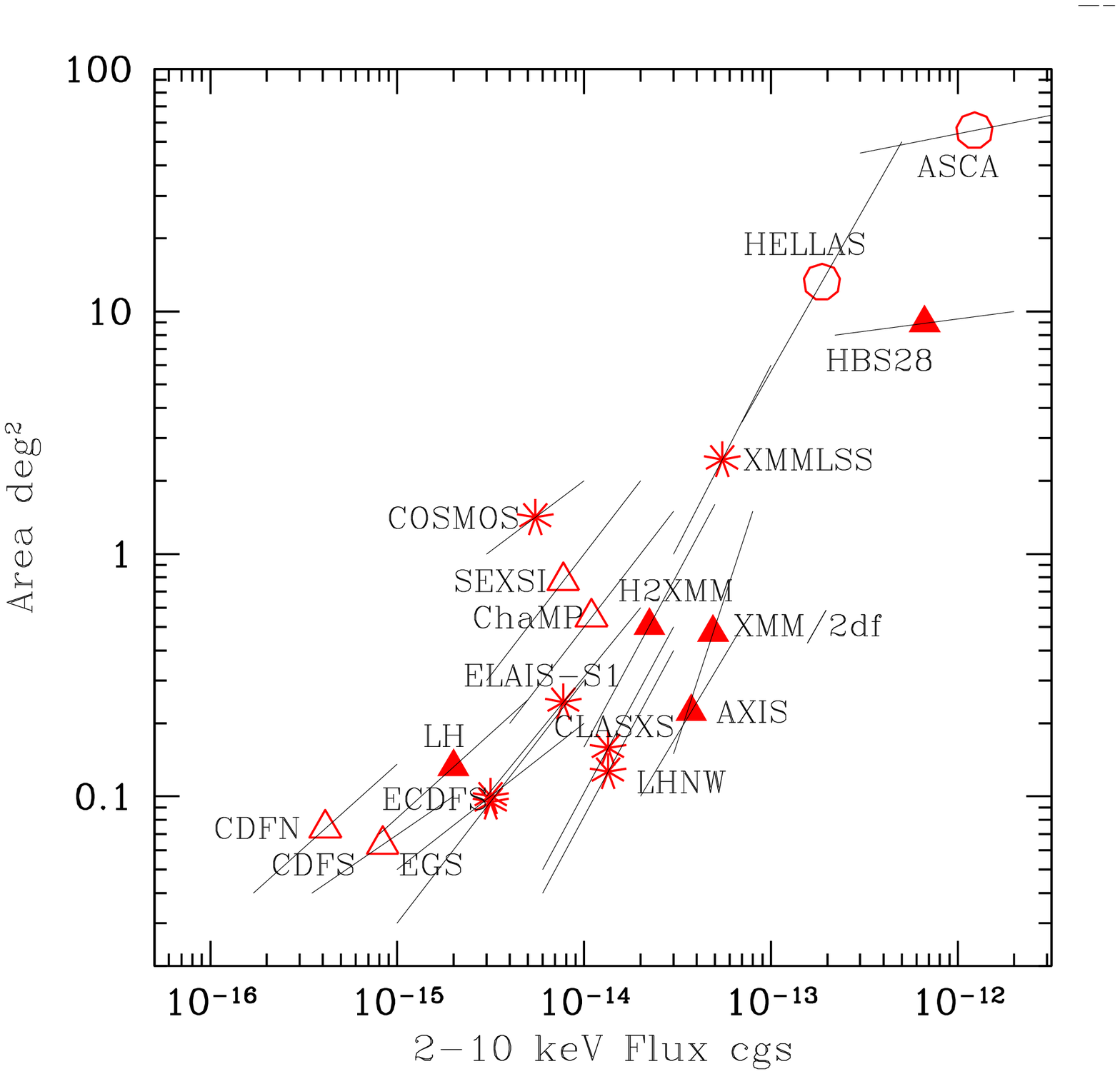}}
\caption{\footnotesize Solid angles and flux limits of AGN surveys carried
out in the 2-10 keV band.
Triangles represent serendipitous surveys constructed from a
collection of pointed observations (Chandra open symbols; XMM-Newton filled
symbols; BeppoSAX and ASCA surveys are also reported). 
The asterisks represent surveys covering contiguous areas.
}
\label{agnsurv}
\end{figure}

Figure \ref{agnsurv} gives an overview of the flux limits and surveyed
areas of major AGN surveys carried out over the last years in the 2-10
keV band.  In this paper we use the source samples given in Table 1,
which include the deepest surveys performed with Chandra as well as
larger area surveys performed with XMM-Newton and ASCA, plus the
so-called Piccinotti sample of local AGN.

We estimated the 2--10 keV luminosity function by fitting the expected
number of AGN in bins of luminosity, redshift and rest frame absorbing
column density N$_H$ (La Franca et al. 2005). This allows us to take
into account observational selection effects. In particular, we
correct for the selection effect due to X-ray absorption, and for the
incompleteness of the optical spectroscopy identification (see La
Franca et al.  2005 for details). Our results extend those of Fiore et
al. (2003), Cowie et al. 2003 and Barger et al. (2005). In all these
three papers no correction for the X-ray absorption is adopted. Fiore
et al. (2003) assign a redshift to the sources without spectroscopic
identification using statistical arguments; Cowie et al.  (2003) and
Barger et al.(2005) estimate upper limits to the AGN density by
assigning to the unidentified sources the redshifts corresponding to
the centers of each L$_X$--z bin.

\begin{table*}[t!]
\caption{\bf 2-10 keV surveys}
{\footnotesize
\begin{tabular}{lcccc}
\hline
Sample     & Tot. Area & Flux limit  & \# sour.  & \% z-spec \\ 
           & deg$^2$   & $10^{-15}$ cgs &        &           \\ 
\hline
HELLAS2XMM       & 1.4   & 6.0       &  232      & 70\%      \\ 
CDFN faint$^a$   & 0.0369& 1.0       &   95      & 59\%      \\ 
CDFN bright$^b$  & 0.0504& 3.0       &   51      & 65\%      \\ 
CDFS faint$^a$   & 0.0369& 1.0       &   75      & 62\%      \\ 
CDFS bright$^b$  & 0.0504& 3.0       &   52      & 60\%      \\ 
Lockman Hole$^c$ & 0.126 & 2.6       &   55      & 75\%      \\ 
HBS28            & 9.8   & 22        &   28      & 100\%     \\ 
AMSSn            & 69    & 30        &   74      & 100\%     \\ 
\hline
Total            &       &           &  662      & 75\%      \\ 
\hline
\end{tabular}

$^a$ Inner 6.5 arcmin radius; $^b$ outer 6.5--10 arcmin annulus; $^c$
inner 12 arcmin radius; $^d$ inner 4.5 arcmin radius. See La Franca et al.
2005 for details.
}
\label{table1} 
\vspace*{-13pt}
\end{table*}

\begin{figure}
\resizebox{\hsize}{!}{\includegraphics[clip=true]{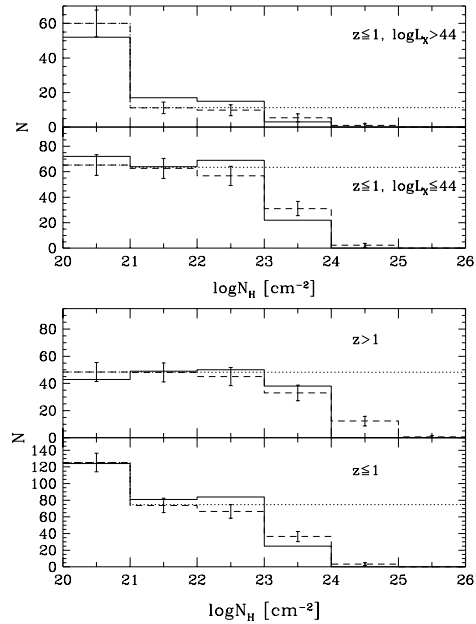}}
\caption{\footnotesize N$_H$ distributions in four luminosity and
redshift bins.  The dotted lines are the assumed  N$_H$
distributions, the dashed lines are the expectations taking into
account the selection effects and the continuous lines are the
observed distributions.  }
\label{nhdist}
\end{figure}

\begin{figure*}[t!]
\begin{tabular}{cc}
\includegraphics[width=6.5cm]{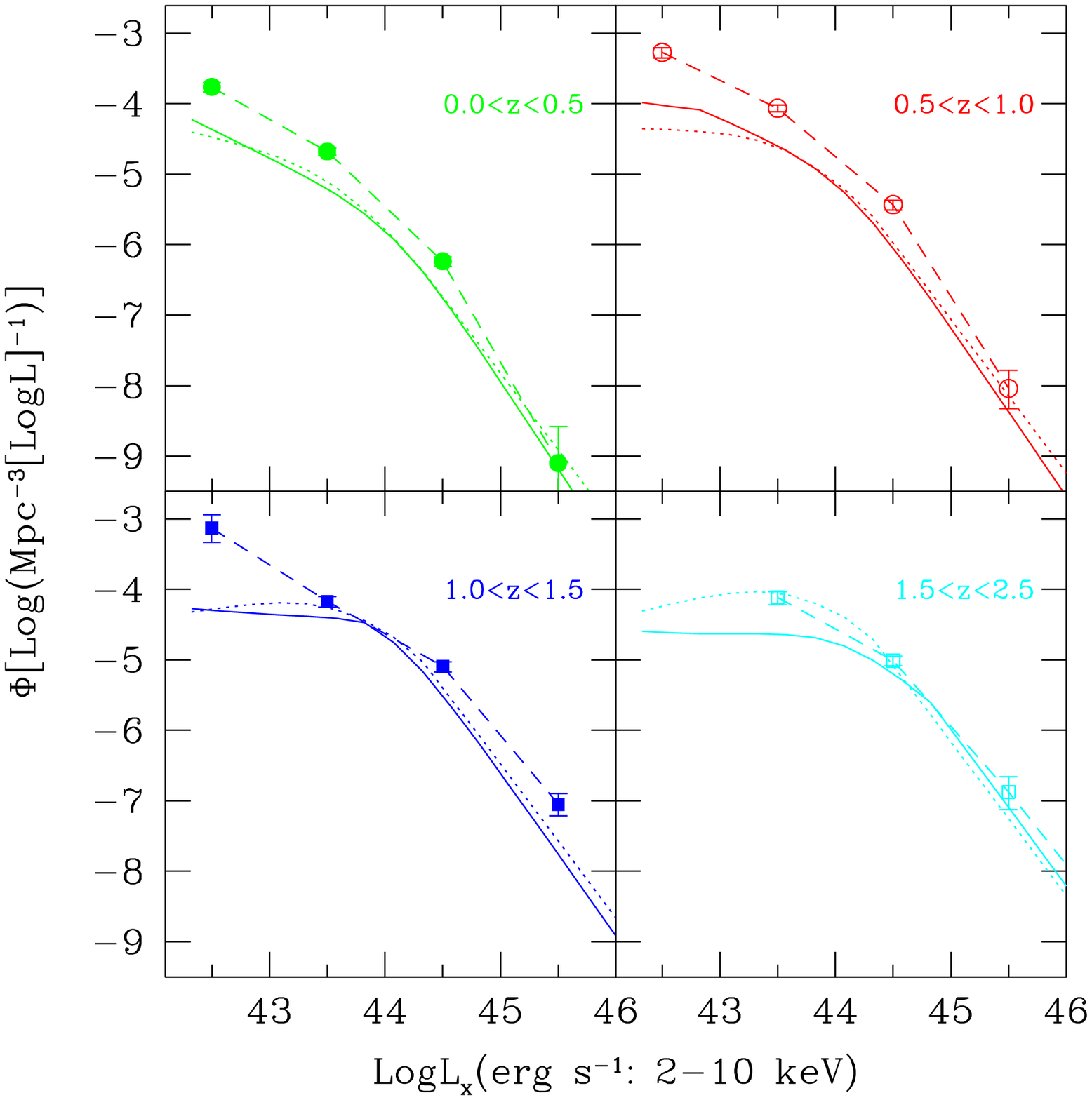}
\includegraphics[width=6.5cm]{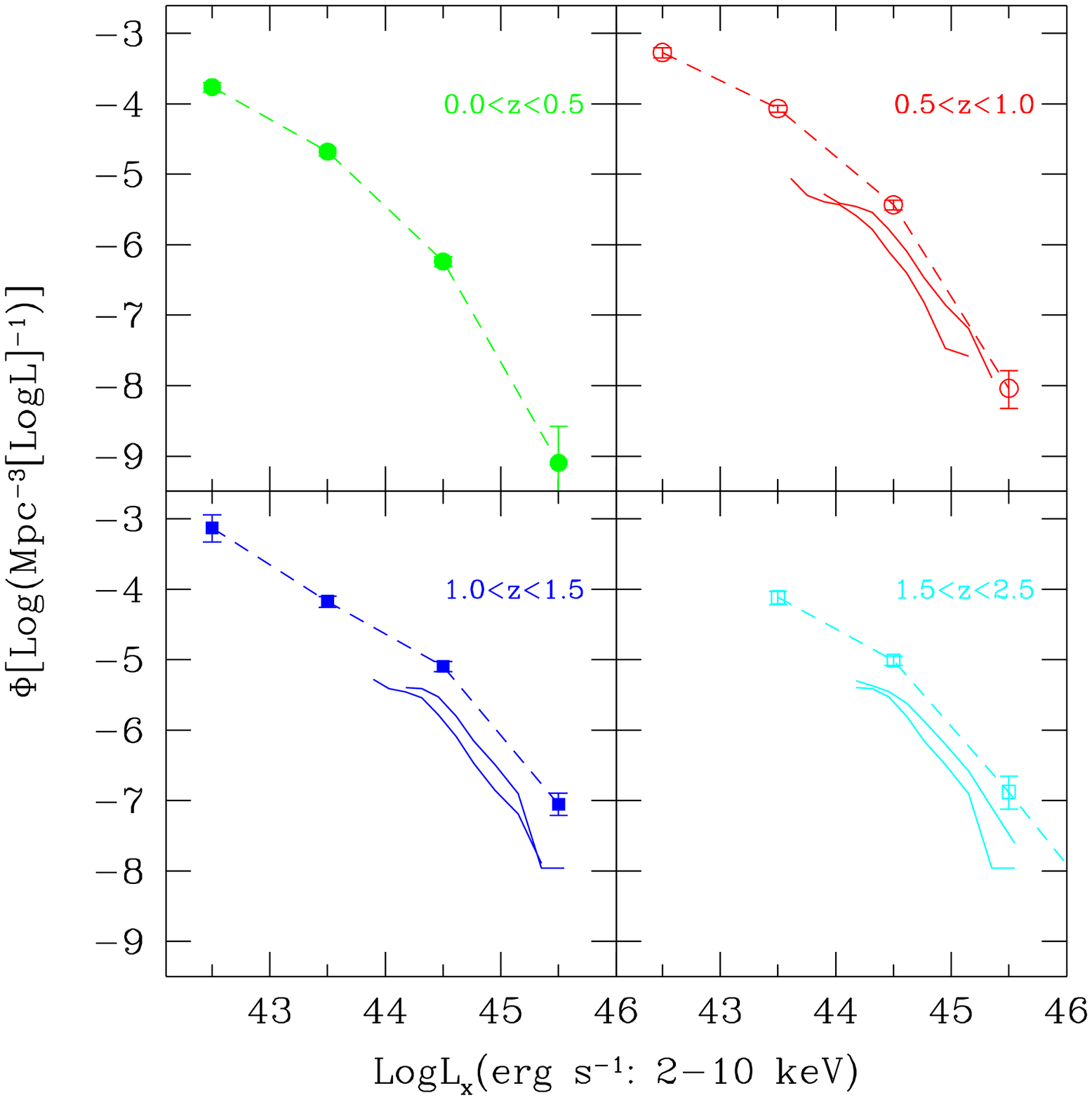}
\end{tabular}
\caption{\footnotesize The 2-10 keV luminosity function in four
redshift bin, compared to the 0.5-2 keV luminosity function of
Miyaji et al (2000, left panel, dotted lines), with the 0.5-2 keV type
1 AGN luminosity function of Hasinger et al. (2005, left panel, solid
lines) and with the optical (B band) luminosity function of Croom et
al. (2004, right panel).  Two optical luminosity function are plotted
in each quadrant, corresponding to redshift bridging the range used
for the 2-10 keV luminosity function.  The conversion factor to pass
from the 0.5-2 keV to the 2-10 keV band has been calculated assuming a
power law spectrum with $\alpha=0.8$ while that to pass from the B
band to the 2-10 keV band has been computed following Marconi et
al. 2004.  }
\label{lf}
\end{figure*}

Figure \ref{nhdist} and \ref{lf} show the best fit N$_H$ distribution
and best fit luminosity functions for a luminosity-dependent density
evolution (LDDE) model (La Franca et al. 2005, see also Miyaji et
al. 2000 for a similar parameterization). The parameters of the
evolving luminosity function and of the dependencies of the N$_H$
distribution by L(2-10 keV) and z have been fitted simultaneously.
The intrinsic N$_H$ distribution (dotted lines in figure \ref{nhdist})
is flat above $10^{21}$ cm$^{-2}$, while the fraction of objects with
N$_H<10^{21}$ cm$^{-2}$ is one of the model parameters.  The dashed
lines show the predictions when the selection effects due to X-ray
absorption, which pushes sources below the X-ray detection limit, and
to the incompleteness of the spectroscopic identification are taken
into account. As expected, most AGN with N$_H>10^{24}$ cm$^{-2}$ are
lost in even the deepest Chandra and XMM 2-10 keV surveys.  In
addition, note that about 2/3 of the AGN with $10^{23}<$N$_H<10^{24}$
cm$^{-2}$ are also lost at z$<1$. Our best fit luminosity function
recovers all these highly X-ray obscured AGN, as well as optically
obscured AGN, whose optical counterpart is too faint to allow a
spectroscopic identification.


The best fit 2-10 keV luminosity function is compared to the best fit
0.5-2 keV total luminosity function of Miyaji et al. (2000) and to the
best fit 0.5-2 keV type 1 AGN luminosity function of Hasinger et
al. (2005) in figure \ref{lf}. These functions falls shorter than the
2-10 keV luminosity function by a factor 3-10 at luminosities
logL$_X=42.5-43$ in the three lowest redshift bins. (At z$>1.5$ the
comparison is less informative because the present 2-10 keV data do
not go deep enough to provide samples of low luminosity AGNs large
enough to constrain adequately their space density.) Conversely, a
better agreement between the 0.5-2 keV and the 2-10 keV luminosity
functions is found at the highest luminosities sampled, with the
exception of the z=1--1.5 bin.  Hasinger et al. (2005) exclude from
their analysis X-ray (and/or) optically obscured AGN, which in any
case are not an important population in 0.5-2 keV surveys. Indeed the
Hasinger et al.  luminosity function is nearly identical to the Miyaji
et al. (2000) one at least at z$<1.5$.  (At z$>1.5$ the Miyaji et
al. 2000 best fit over-predicts the number of low luminosity AGN,
probably because to the larger uncertainties at high redshift and low
luminosity due to the much shallower data used in comparison with
Hasinger et al. 2005).  Obscured AGNs are recovered in the 2-10 keV
luminosity function. The result is that the evolution of this
luminosity function deviates less from a pure luminosity evolution
than the soft X-ray luminosity functions. At least part of the strong
luminosity dependent density evolution claimed based on the 0.5-2 keV
data is therefore due to the exclusion of obscured AGN.

Figure \ref{lf} compares the 2-10 keV and optical luminosity
functions. Luminosity dependent conversion factors to pass from the B
band to the 2-10 keV band have been computed following Marconi et
al. (2004). It should be borne in mind that the uncertainty on the
conversion factor may be as large as a factor of two.  Note that the
low luminosity end of the optical luminosity function is always 1, 1.5
dex higher than that of the X-ray luminosity function. This is due to
the difficulty in selecting low luminosity AGNs against their host
galaxy in the optical band.  The optical luminosity function is
similar or slightly lower than the 0.5-2 keV luminosity function at
z$>0.5$, and therefore the same comments given above for the soft
X-ray luminosity function apply also in this case: e.g. a large
fraction of obscured accretion may be missed by both optical and soft
X-ray selection.

\begin{figure}
\includegraphics[width=6.5cm]{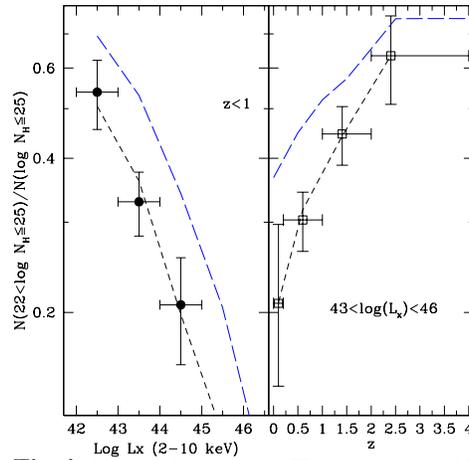}
\caption{\footnotesize Observed fraction of X-ray obscured
(N$_H>10^{22}$ cm$^{-2}$) AGN as a function of L(2-10keV) and z.  The
long dashed lines are the best fit intrinsic distributions.  The short
dashed lines are the expectations taking into account all selection
effects.  }
\label{nhfrac}
\end{figure}

Figure \ref{nhfrac} shows the fraction of X-ray obscured AGN
(N$_H>10^{22}$ cm$^{-2}$) as a function of the 2-10 keV luminosity and
the redshift.  The long dashed lines are the best fit intrinsic
distributions while the short dashed lines are the expectations taking
into account all selection effects described above.  Both the observed
and best fit fraction of obscured AGN at z$<$1 decrease strongly with
the AGN luminosity, a behavior already noticed in the literature
since the first Einstein systematic observations of QSO (Lawrence \&
Elvis 1982), and confirmed quantitatively by Ueda et al. (2003). Note
that this trend becomes more evident putting together deep surveys,
well sampling low luminosity AGN, and large area surveys, which can
provide large samples of luminous AGN. Note also that selection
effects affects in a similar way low and high luminosity AGN: the
ratio of the intrinsic and predicted fractions of obscured AGN is
nearly constant with the luminosity. This is not the case when
considering the dependence of the fraction of obscured AGN with the
redshift. The reason is that obscured AGN are much more likely to be
missed in X-ray surveys at low redshift than at high redshift, because
the photoelectric cut-off is quickly redshifted toward low X-ray
energies.  Indeed, comparing the best fit intrinsic distribution to
the distribution expected after the selection effects, we note that
while we loose just about 15\% of the obscured AGN at z=2 slightly
less than half are lost at z$<0.2$. By correcting for this selection
effect we find that the intrinsic fraction of obscured AGN in the
luminosity range $10^{43}-10^{46}$ still increases by a factor 1.8
from z=0 to z=2.  This new result is due a better coverage of the
luminosity-redshift diagram in comparison with previous work.

\begin{figure*}[t!]
\resizebox{\hsize}{!}{
\begin{tabular}{cc}
\includegraphics[clip=true]{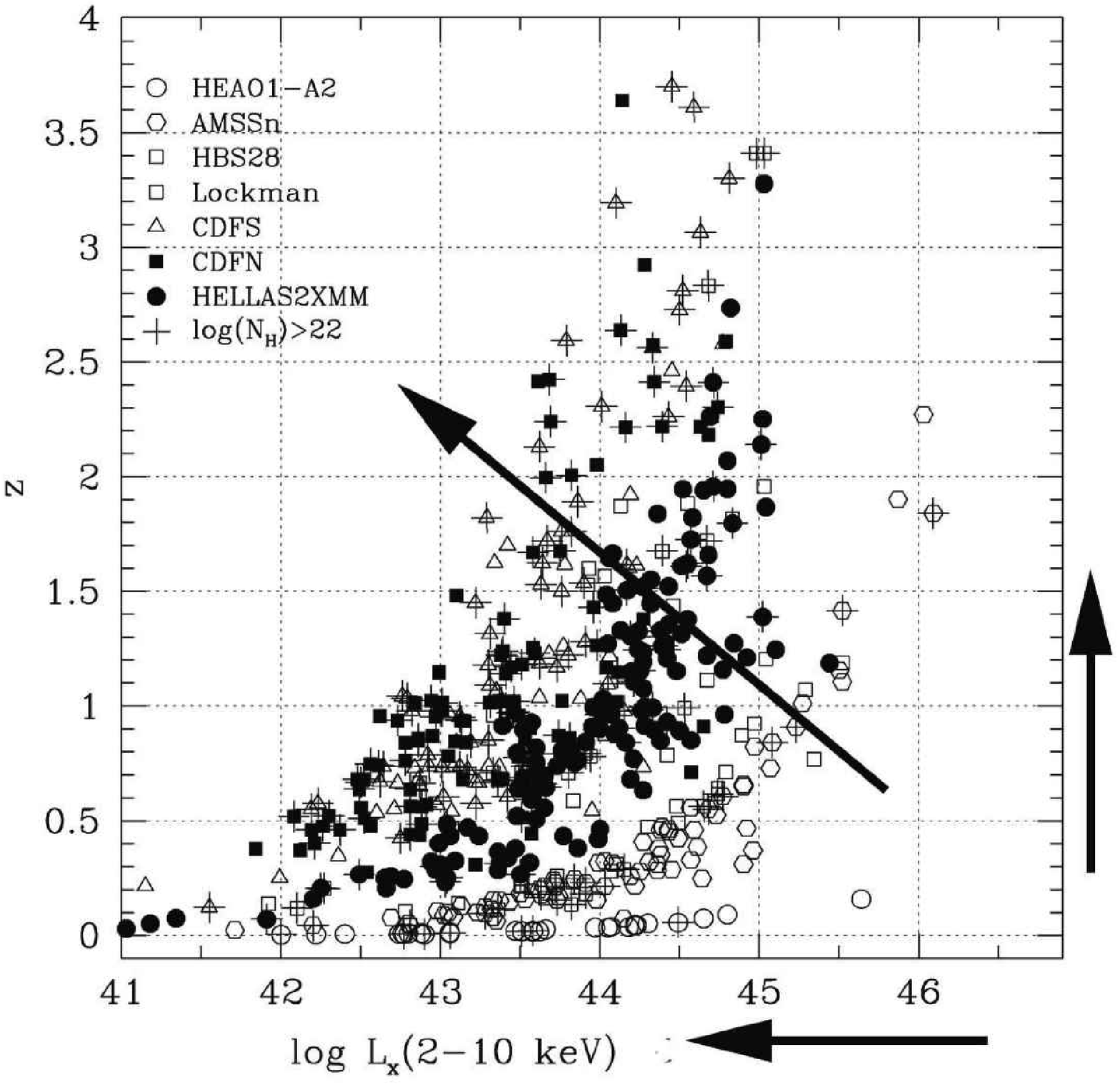}
\includegraphics[clip=true]{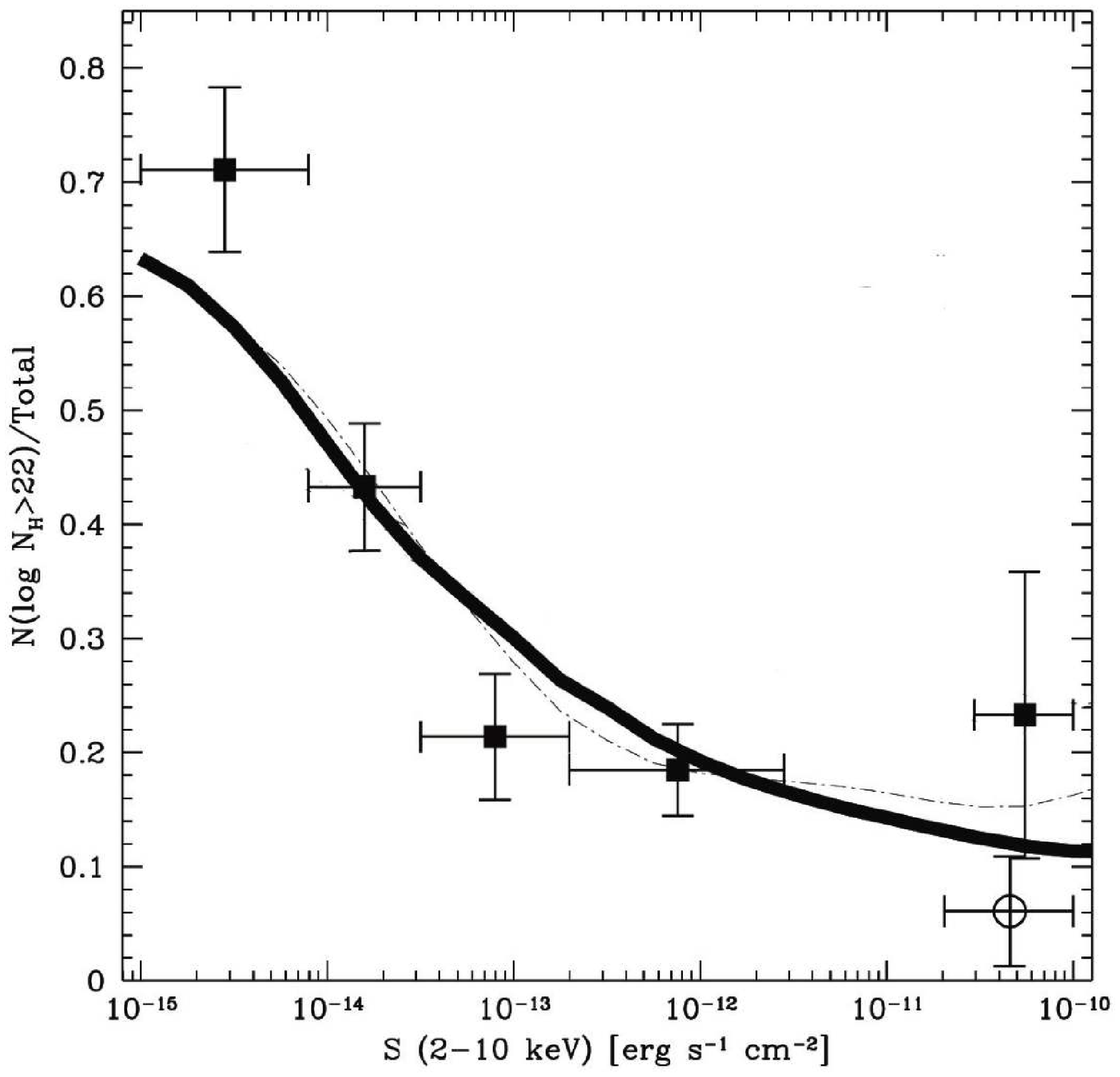}
\end{tabular}
}
\caption{\footnotesize Left panel: the luminosity-redshift plane for
the surveys in Table 1.  Crosses indicate AGN with N$_H>10^{22}$ 
cm$^{-2}$. 
The arrows indicate the directions in which
the fraction of obscured AGN increases, according to figure
\ref{nhfrac}.  Right panel: observed fraction of X-ray obscured
(N$_H>10^{22}$ cm$^{-2}$) AGN as a function of F(2-10keV) }
\label{lxz}
\end{figure*}

Figure \ref{lxz} shows the luminosity-redshift plane for the
surveys in Table 1.  The arrows indicate the directions in which the
fraction of obscured AGN increases, according to figure
\ref{nhfrac}. Excluding the CDFS and CDFN data would strongly limit
the number of AGN with logL(2-10)$<44$ (those more likely to be
obscured, according to the left panel of figure \ref{nhfrac}), making
difficult to asses any trend with the redshift.  The arrows indicate
also the direction in which the flux decreases. Indeed a strong
correlation of the number of obscured AGN and the flux has been
reported in the past (Piconcelli et al. 2003, Ueda et al. 2003, Perola
et al. 2004) and it is expected by AGN synthesis models of the CXB
(Comastri et al. 2001). Figure \ref{lxz} shows this correlation for the
source sample in Table 1, along with the best fit LDDE model of La
Franca et al. (2005).  In conclusion, the correlations found between
the fraction of obscured AGN and luminosity and redshift of figure
\ref{nhfrac} confirm and extend previous determinations, based on
smaller and shallower surveys.

\section{Discussion}

\begin{figure}
\resizebox{\hsize}{!}{\includegraphics[clip=true]{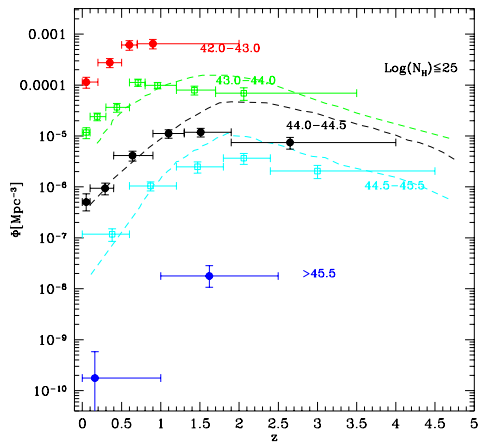}}
\caption{\footnotesize 
 The evolution of the number density of AGNs selected in the 2-10 keV band 
in three bins of luminosity  
($43<log L_X <44$ $44<log L_X<44.5$   
$44.5<log L_X$ compared to the prediction of the model of Menci et al. (2004)}
\label{menci}
\end{figure}

Our determination of the 2-10 keV AGN luminosity function, accounting
for selection effects due to nuclear obscuration by gas and dust,
confirms the AGN differential luminosity evolution, but makes it less
extreme than what is found selecting unobscured AGN only (see
e.g. Hasinger et al. 2005). This is important for both models that
make use of the AGN luminosity function (to reproduce the X-ray and IR
Cosmic backgrounds for example), and for models which try to explain
the AGN luminosity function, as the semi-analytic, hierarchical
clustering model proposed by Menci et al. (2004).  Figure \ref{menci}
compares the AGN number density as a function of z with the
predictions of the Menci model.  The result is qualitatively similar
to that reported by Menci et al. (2004). The trend of lower luminosity
AGN peaking and increasingly lower redshift is found in both data and
model, which however predicts a number of low-to-intermediate
luminosity (Seyfert like) AGN at z=1.5-2.5 higher than what is
observed by a factor of a few.  This disagreement can be due to at
least two broad reasons: either La Franca et al. (2005) underestimate
the number of highly obscured AGN missed at z=1.5-2.5 in Chandra and
XMM-Newton surveys, or there are problems with the model.

About the first possibility, it must be noted that most obscured AGN
selected below 10 keV have column densities in the range $N_H \sim
10^{22-23}$ cm$^{-2}$ with only a handful of the faintest sources
which may be Compton thick ($N_H > 10^{24}$ cm$^{-2}$, just $\sim4\%$
in the CDFS, Tozzi et al. 2006). So we still may be viewing just the
tip of the iceberg of highly obscured sources. Our
estimates of the number of obscured AGN missed in today X-ray surveys
are based on large extrapolations from what we know about the fraction
of obscured AGN in the local Universe. Compton thick objects may well
be more common at high redshift, as suggested, for example, by Fabian
(1999) and Silk \& Rees (1998).  An alternative approach to find
Compton thick AGN at z$>$1 is to select sources with AGN luminosities
in the mid--infrared and faint near--infrared and optical emission
(Martinez-Sansigre et al. 2005).  These authors estimate that probably
more than half of high luminosity QSOs are highly obscured, although
with quite large uncertainties. Unfortunately, the X--ray properties
of these infrared selected sources are not known, and therefore it is
difficult to understand how the mid--infrared selection compares with
the X--ray one. In particular, it is not clear which is the fraction
of the mid--infrared selected type 2 AGN which would have been
selected by X--ray surveys. Answers to these questions will soon come
from the study of fields with both X--ray and mid--infrared coverage
(e.g. the ELAIS-S1 field, Puccetti et al. 2006, Feruglio et al. 2006
in preparation, and the COSMOS fields), and from deep X--ray follow-up
observations of the mid--infrared selected sources in the Spitzer
First Look Survey.  
\begin{figure*}[t]
\resizebox{\hsize}{!}{\includegraphics[clip=true]{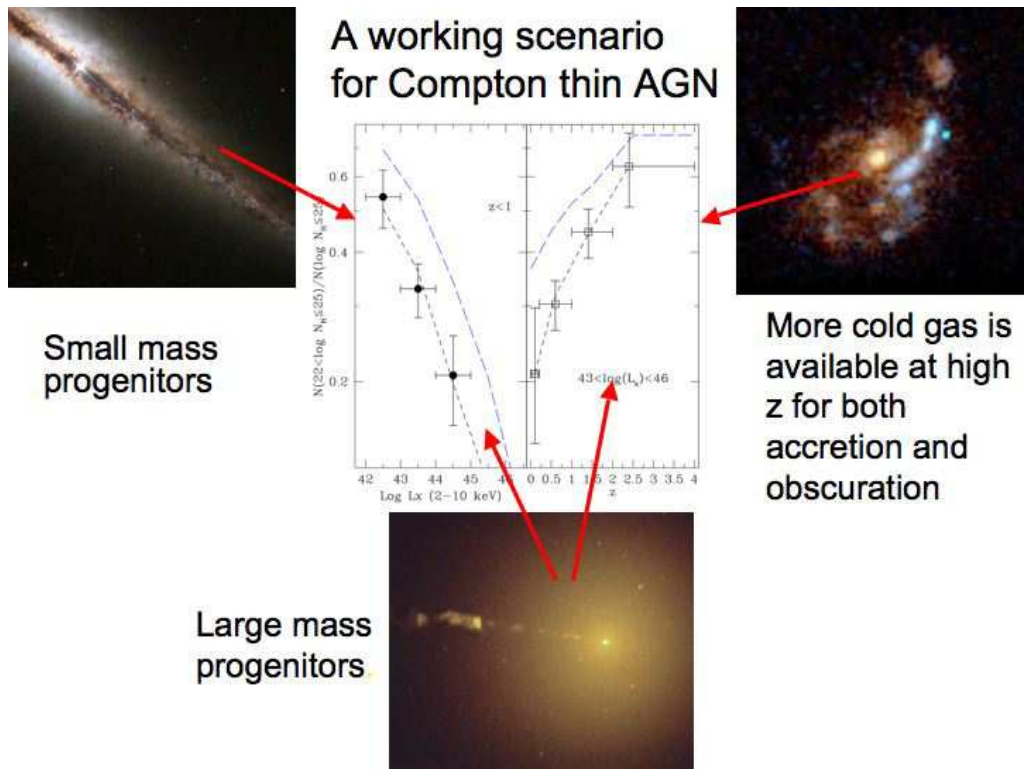}}
\caption{\footnotesize 
A working scenario for Compton thin AGN}
\label{cartoon}
\end{figure*}

To avoid any possible selection effect, for an unbiased census of the
AGN population making the bulk of the CXB and an unbiased measure of
the AGN luminosity function at z=1--2, sensitive observations
extending at the peak of the CXB are clearly needed.  More
specifically to resolve $\sim50\%$ of the CXB in the 20--40 keV band
we need to go down to fluxes of $10^{-14}$ \cgs in this band.  This
can be achieved only by imaging X-ray telescopes, (see e.g. Fiore et
al. 2004, Ferrando et al. 2005).  Key issues are: a) high collecting
area; b) sharp PSF (15 arcsec or less Half Energy Diameter); c)
low detector internal background.

About models, as in all complex processes, there are several areas
which may be critical. For example, the prescriptions adopted by the
Menci model to switch an AGN on and to compute its feedback on the
host galaxy may be too simple, or, more in general, the descriptions
of the mechanisms regulating the amount of cool gas in low-mass host
galaxies and the physical mechanism at work at small accretion rates
may be inadequate, as well as the statistics of DM condensations.
Further constraints to this model, which may shed light on these
issues, come from the observed correlations between the fraction of
obscured AGN with luminosity and redshift. These correlations are at
odds with popular AGN unified Schemes (see e.g. Lamastra et al. 2006),
and may suggest that low luminosity, Seyfert like AGN and powerful
QSOs are intrinsically different populations, with different
obscuration properties, caused by different formation histories, a
bimodal behaviors reminding that of the color distribution of galaxies
(see e.g. Menci et al. 2005).  The Seyfer-like object population could
be due to nuclear activation during loose galaxy encounters (fly-by,
Cavaliere \& Vittorini 2000) at sub Eddington levels, and the second
population could be due to nuclear activation during major mergers, in
the process of galaxy assembly.  Seyfert like AGN could be mostly
associated to galaxies with merging histories characterized by small
mass progenitors while QSOs may be associated to large mass
progenitors, as sketched in the cartoon of figure \ref{cartoon}. In
the first case nuclear accretion and star-formation could be
self-regulated by feedbacks, which can therefore be effective in
leaving available cold gas that can both cause an obscuration of the
nucleus and be accreted during subsequent galaxy encounters. In these
galaxies gas and dust lanes can efficiently obscure the nucleus along
many lines of sight (a scenario similar to that outlined by Matt
2000).  In the second case case feedback could be less effective in
reheating/expelling the cold gas, most of which is rapidly converted
in stars at high z.  The obscuration properties of the two populations
could be different in terms of gas geometry, covering factor, density,
ionization state, metallicity, dust content and composition. A
quantitative comparison between the prediction of the Menci model
about the fraction of obscured AGNs with the observation is in
progress. This comparison can help in both understanding which is the
leading physical mechanism responsible for the activation and
obscuration of AGNs of different luminosities, and in understanding
the role of relative feedbacks between nuclear activity, star
formation and galaxy evolution, as a function of the host mass and
luminosity.

\begin{acknowledgements}
The original matter presented in this paper is the result of the
effort of a large number of people, in particular of the {\tt
HELLAS2XMM} collaboration: A. Baldi, M. Brusa, N. Carangelo,
P. Ciliegi, F. Cocchia, A. Comastri, V. D'Elia, C. Feruglio, F. La
Franca, R. Maiolino, G. Matt, M. Mignoli, S. Molendi, G. C. Perola,
S. Puccetti N. Sacchi, and C.Vignali. I wish also to thank M. Elvis,
P. Severgnini, N. Menci, A. Cavaliere, R. Gilli, G. Pareschi and
O. Citterio.  This work was partially supported by the Italian Space
Agency (ASI) under grant I/023/05, by INAF under grant \# 270/2003
and MIUR grant Cofin--03--02--23.
\end{acknowledgements}

\bibliographystyle{aa}

\end{document}